\documentclass[preprint2,twoside]{hwo}

\usepackage{graphicx}
\usepackage{tabularx,ragged2e}
\newcolumntype{C}{>{\Centering\arraybackslash}X} 

\input{hwo.h}

\begin{document}

\title{\textbf{\LARGE Detecting and characterising the magnetic field of exoplanets}}
\author {\textbf{\large A. Strugarek$^{1}$, S. V. Berdyugina$^2$, V. Bourrier$^3$, J. A. Caballero$^4$, J. J. Chebly$^1$, R. Fares$^5$, A. Fludra$^6$, L. Fossati$^7$, A. García Muñoz$^1$, L. Gkouvelis$^8$, C. Gourvès$^1$, J. L. Grenfell$^9$, R. D. Kavanagh$^{10}$, K. G. Kislyakova$^{11}$, L. Lamy$^{16,12}$, A. F. Lanza$^{13}$, C. Moutou$^{14}$, D. Nandy$^{15}$, C. Neiner$^{16}$, A. Oklopčić$^{17}$, A. Paul$^1$, V. Réville$^{14}$, D. Rodgers-Lee$^{18}$, E. L. Shkolnik$^{19}$, J. D. Turner$^{20}$, A. A. Vidotto$^{21}$, F. Yang$^1$, P. Zarka$^{16}$ }}

\affil{$^1$\small\it Université Paris-Saclay, Université Paris Cité, CEA, CNRS, AIM, 91191,Gif-sur-Yvette, France}
\affil{$^2$\small\it Istituto ricerche solari Aldo e Cele Daccó (IRSOL), Universitá Svizzera italiana, Locarno, Switzerland}
\affil{$^3$\small\it Département d'Astronomie de l'Université de Genève, Switzerland}
\affil{$^4$\small\it Centro de Astrobiología CSIC-INTA, 28692 Villanueva de la Cañada, Madrid, Spain}
\affil{$^5$\small\it  Department of Physics, College of Science, United Arab Emirates University, PO Box No. 15551, Al Ain, UAE}
\affil{$^6$\small\it UKRI/STFC, RAL Space, OX11 0QX, UK}
\affil{$^7$\small\it Space Research Institute, Austrian Academy of Sciences, Schmiedlstrasse 6, 8042, Graz, Austria}
\affil{$^8$\small\it Universitäts-Sternwarte, Ludwig-Maximilians-Universität München, Scheinerstrasse 1,München, 81679, Germany}
\affil{$^9$\small\it Department of Extrasolar Planets and Atmospheres (EPA), German Aerospace Centre (DLR), Rutherford  str. 2, 12489 Berlin, Germany}
\affil{$^{10}$\small\it ASTRON, The Netherlands Institute for Radio Astronomy, Oude Hoogeveensedijk 4, 7991 PD Dwingeloo, The Netherlands}
\affil{$^{11}$\small\it Department of Astrophysics, University of Vienna, Austria}
\affil{$^{12}$\small\it LAM, CNRS, Pythéas, France}
\affil{$^{13}$\small\it INAF-Osservatorio Astrofisico di Catania, Via S. Sofia, 78 - 95123 Catania, Italy}
\affil{$^{14}$\small\it IRAP, CNRS, Université de Toulouse}
\affil{$^{15}$\small\it Center of Excellence in Space Sciences India, Indian Institute of Science Education and Research Kolkata, Mohanpur 741246, India}
\affil{$^{16}$\small\it LIRA, Paris Observatory, CNRS, PSL University, Université Paris Cité, Sorbonne University, CY Cergy University, 5 place Jules Janssen, 92195 Meudon, France}
\affil{$^{17}$\small\it University of Amsterdam, Netherlands}
\affil{$^{18}$\small\it Astronomy \& Astrophysics Section, School of Cosmic Physics, Dublin Institute for Advanced Studies, 31 Fitzwilliam Place, Dublin D02 XF86, Ireland}
\affil{$^{19}$\small\it School of Earth and Space Exploration, Arizona State University, USA}
\affil{$^{20}$\small\it Department of Astronomy and Carl Sagan Institute, Cornell University, Ithaca, NY, USA}
\affil{$^{21}$\small\it Leiden Observatory, Leiden University, the Netherlands}

\author{\footnotesize{\bf Endorsed by:}
James Barron (Queen's University), Jean-Claude Bouret (Laboratoire d'Astrophysique de Marseille), Diego Godoy-Rivera (Instituto de Astrofísica de Canarias), Ana Ines Gomes de Castro (Universidad Complutense de Madrid), Jasmina Lazendic-Galloway (Eindhoven University of Technology), Eunjeong Lee (EisKosmos (CROASAEN), Inc.), Evelyn Macdonald (University of Vienna ), Ignacio Mendigutia (Centro de Astrobiologia (CAB, CSIC-INTA)), David Montes (UCM, Universidad Complutense de Madrid), Faraz Nasir Saleem (Egypt Space Agency), Gaetano Scandariato (INAF), Jessica Spake (Carnegie Observatories), Daniel Valentine (University of Bristol), Victor Vincent (University College Dublin), Peter Wheatley (University of Warwick, UK), Siayo Xu (University of Florida).
}

\begin{abstract}
Magnetic fields play a crucial role in planetary evolution and habitability. While the intrinsic magnetic fields of solar system planets are relatively well understood, the magnetic properties of exoplanets remain largely unconstrained, despite their potential ubiquity. Detecting exoplanetary magnetic fields is essential to advancing our understanding of planetary habitability beyond the solar system. This paper focuses on two promising spectropolarimetric techniques for detecting magnetic fields in hot exoplanets: direct detection through polarization signatures in the He I 1083 nm triplet and indirect detection via star-planet magnetic interactions manifesting as stellar hot spots. The direct method is particularly suited to close-in gas giants, leveraging the Hanle and Zeeman effects to detect low-amplitude magnetic fields. The indirect method can apply to both giant and low-mass planets by identifying magnetic connectivity-induced features in the stellar atmosphere. Although the interpretation of current detections remain tentative, upcoming high-resolution spectropolarimetric capabilities in the UV and near-infrared, particularly with future missions like HWO, promise to enable definitive measurements of exoplanetary magnetic fields. These advancements will open new avenues for probing the magnetic environments of exoplanets and their implications for atmospheric retention and habitability.
\end{abstract}

\vspace{2cm}

\section{Science Goal}

{\bf What kind of magnetic field do hot exoplanets harbor?}
 
On Earth, it is thought today that the magnetosphere plays a paramount role in protecting our atmosphere from erosion by the ambient solar wind and protects us from high-energy particles. The intrinsic magnetic field of planets in the solar system are quite diverse, with sometimes even only a weak remnant field, as on Mars. For exoplanets, however, we are today nearly blind to their magnetic characteristic, which is of paramount importance to the very concept of planetary habitability for low-mass planets, as well as obtaining indirect constraints on their internal structure. So far we have no strong constraints on the occurrence rate of magnetospheres, although they are likely ubiquitous for exoplanets. Therefore any firm detection would provide a significant step forward in understanding how common exoplanetary magnetic fields are.
 
Two main routes to detect exoplanetary magnetic fields can be identified: direct and indirect detections \citep{brain_exoplanet_2024}. Direct detection methods rely on signals being affected locally by the presence of the exoplanetary field, and could theoretically be obtained in UV, infrared, and in radio. One iconic example is the prediction of polarization of the helium absorption triplet when the stellar photons are transmitted through the atmosphere of the exoplanet \citep{oklopcic_detecting_2020}. In addition, hot exoplanets orbit so close to their star that they can be magnetically connected: depending on the size and shape of their magnetosphere, they can induce detectable hot spots in the atmosphere of the star itself \citep{Cuntz2000,Shkolnik2003,castro-gonzalez_signs_2024}. Upon detection of these features, we can have an indirect access to the properties of the exoplanetary magnetic field \citep{Cauley2019,Strugarek2022,vidotto_space_2023}. Other methods to detect magnetic fields have also been proposed \citep{brain_exoplanet_2024}, e.g. leveraging the existence of bow-shocks \citep{Vidotto2017a}, studies of Ly-alpha transit signatures of exoplanets \citep{Kislyakova2014}, scattered light polarization variations in molecular bands due to exoplanetary magnetic field, i.e., molecular Hanle effect \citep{berdyugina_polarized_2016}, detection of heavy ion escape \citep{savel_new_2024}, and from transit absorption of neutral hydrogen and heavy ions \citep{Ben-Jaffel2022}. In this document, we focus on the two methods that leverage spectropolarimetry.
 
The direct detection of an exoplanetary field has only been tentative so far (e.g. in radio, see \citealt{turner_search_2021}), and the indirect detection of hot spots has been only tentatively associated with star-planet magnetic interactions \citep{strugarek_introduction_2025}. Both detection methods are currently being used in the radio domain, and significant advances are expected in the near future with the SKA \citep{zarka_magnetospheric_2015,callingham_radio_2024}. HWO has the potential to put firm grounds on their detection and interpretation thanks to high-resolution spectropolarimetry in near infra-red (direct detection) and UV (both direct and indirect detection), opening up a whole new chapter for the characterization of exoplanets.

\section{Science Objective}

With an ever increasing number of exoplanets being detected, now numbering over 5800, it is expected that several of them would have magnetic fields ranging from subGauss to hundreds of Gauss strengths (e.g. \citealt{yadav_estimating_2017,McIntyre2019,hori_linkage_2021,kilmetis_magnetic_2024}). So far, though, radio observations of brown dwarfs question the broad applicability of existing dynamo scaling-laws to assess the magnetic field of low-mass objects \citep{Kao2016,kavanagh_unravelling_2024}. For planets, on the one hand, observations and simulations in solar system objects such as Earth, Mars and other (e.g. \citealt{bharati_das_modeling_2019,basak_modelling_2021,paul_global-mhd_2023}) indicate that the interactions of stellar magnetism with planets, with and without magnetospheres, play an important role in atmospheric evolution. On the other hand, simulations of star-planet interactions with varying stellar wind, stellar coronal mass ejection and planetary magnetic fields suggest that magnetism critically influences exoplanetary atmospheric mass loss, and therefore habitability \citep{gupta_impact_2023,presa_atmospheric_2024,hazra_magnetic_2025}.

We propose two objectives to obtain a firm detection of an exoplanetary field, leveraging direct and indirect detection methods (Brain et al. 2024) and relying both on high-resolution spectropolarimetry. The first method generally targets giant planets, whereas the second method can equally aim for giant and low-mass planets.

\begin{figure*}[ht!]
    \centering
    \includegraphics[width=\textwidth]{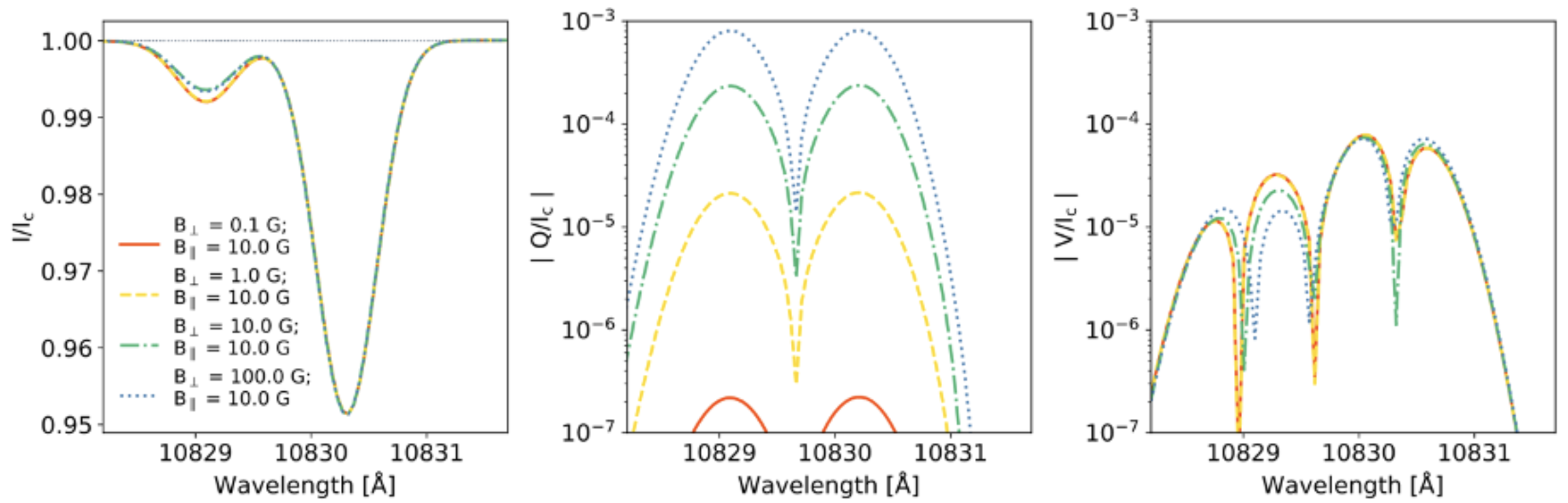}
    \caption{Intensity (left), linear (middle) and circular (right) polarization profile of the He I 1083 nm absorption in a transit of a hot Jupiter. Each color corresponds to different values of line-of-sight ($\parallel$) and perpendicular ($\perp$) magnetic field strengths at the location where He triplet forms (from \citealt{oklopcic_detecting_2020}). The middle panel strikingly shows the sensitivity of the linear polarization (Hanle effect) to the local magnetic field perpendicular to the line-of-sight.}
    \label{fig:fig1}
\end{figure*} 

\subsection{Spectropolarimetric direct detection of an exoplanet magnetic field}

Recent efforts have been carried out to assess He I 1083 nm transmission spectroscopy \citep{oklopcic_new_2018,oklopcic_helium_2019,schreyer_using_2024} through an escaping atmosphere of an exoplanet. \citet{oklopcic_detecting_2020} unveiled that the He I 1083 nm absorption should also present distinct polarimetric signatures tied to the existence of a local magnetic field. 
 
The very presence of neutral helium atoms, particularly in the metastable state, is not guaranteed in the atmosphere of exoplanets, but in the case of close-in gas giants stellar irradiation can generally lead to a significant population of metastable neutral He (but see \citealt{poppenhaeger_helium_2022,chebly_exploring_2025}, rendering the absorption detection possible in tens of cases (e.g. \citealt{spake_helium_2018}), particularly leveraging high spectral resolution (e.g. \citealt{orell-miquel_mopys_2024}). In this state, the atoms absorb the linear polarization of the incoming stellar light due to the Hanle effect and the circular polarization due to the Zeeman effect. The former strongly depends on the magnetic field perpendicular to the line-of-sight (Figure \ref{fig:fig1}).  
Provided enough sensitivity is obtained on the linear polarization, the strength of this method lies in its ability to probe low amplitude magnetic fields at the location He I 1083 nm triplet forms, theoretically down to a few $10^{-4}$ G \citep{oklopcic_detecting_2020}. It is also sensitive to the (mis)alignment of the magnetic field with the line of sight, and should favorably be detectable for planets around relatively bright stars and presenting large transit depths. In practice, direct detection thanks to high precision spectropolarimetry of He I 1083 nm triplet will be primarily feasible for close-in gas giants around solar-like stars, and for moderate magnetic field strengths of about 10 G and more ($Q$/$I_c$ $>$ $10^{-4}$, see Figure \ref{fig:fig1}). 

\begin{figure*}[ht!]
    \centering
    \includegraphics[width=\textwidth]{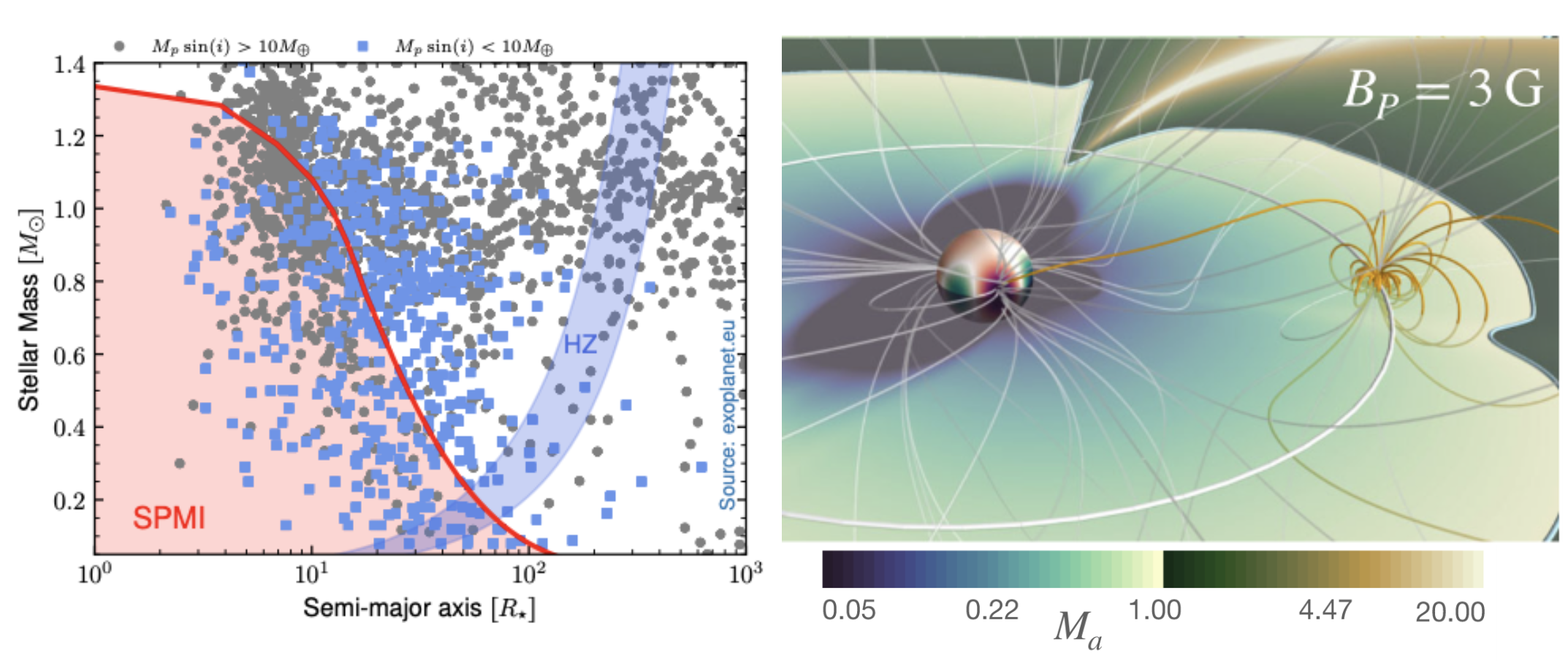}
    \caption{Left: The known exoplanet population (divided into two bins of masses above and below 10 Earth masses) as a function of their semi-major axis and of the central stellar mass. The classical habitable zone is shown by the blue area, and the region where star-planet magnetic interactions (SPMI) are expected to occur is shown by the red area, i.e. within the so-called Alfvén surface. Right: Modelling of the wind of HD 189733 based on a spectropolarimetric map \citep{Fares2017a} showing the magnetic connectivity between the hot Jupiter HD 189733 b and its host (from \citealt{Strugarek2022}).}
    \label{fig:fig2}
\end{figure*}

\subsection{Spectropolarimetric indirect detection of an exoplanet magnetic field
}

The indirect characterization of the magnetic field of a planet thanks to star-planet magnetic interaction has been a grand goal for more than two decades \citep{Cuntz2000}. It is a notoriously difficult signal to detect for multiple reasons. First, it requires the planet to orbit sufficiently close to its star to allow for the magnetic connection to occur \citep{Lanza2009a}. As shown in the left panel of Figure \ref{fig:fig2}, a large portion of the hot exoplanets are actually likely orbiting within the star-planet magnetic interaction limit (red zone). Second, the hot spots created by the interaction add to an already complex stellar atmosphere due to the star’s magnetic activity \citep{Ip2004,paul_stellar_2025}, and the two signals can be difficult to disentangle \citep{Lanza2012a}. Third, it is modulated by the orbit of the planet, the rotation rate of the star, and the structure of the magnetic field connecting the planet to the star \citep{strugarek_interactions_2023}. It is therefore expected that the signal is subject to on/off situations on timescales varying from days (orbit) to years (stellar activity modulation; \citealt{Shkolnik2003}). Planets on eccentric orbits are also a prime target to monitor for such on/off effects \citep{maggio_coordinated_2015,figueira_is_2016,castro-gonzalez_signs_2024}. We know, nevertheless, that this mechanism is robust since the same phenomenon excites UV (and radio) bright spots at the magnetic poles of Jupiter due to the orbit of its natural satellites \citep{zarka_planet_2007,Saur2013}. Furthermore, this detection method is agnostic to the exoplanetary type: it only requires an exoplanetary magnetosphere to exist in the first place.
 
For these reasons, only tentative characterisation of the star-planet magnetic interaction signal have been reported so far (e.g. \citealt{Cauley2019}). The advances in modelling of the environment of cool stars (propelled by advances in the solar corona and wind modelling, see e.g. \citealt{macneice_assessing_2018}) coupled to the future capabilities of HWO (broad UV coverage coupled with large collecting area) now render the firm confirmation of such a tentative characterisation possible. 
 
The firm interpretation of stellar hot spots as originating from star-planet magnetic interactions requires the knowledge of the large-scale magnetic connectivity between the orbit of the planet and the star, and of the thermodynamic structure of the atmosphere. High resolution UV-Visible-IR spectro-polarimetry would provide unprecedented inference of both the magnetic and thermal structure of the upper atmosphere of the star. Leveraging such high-resolution observations covering a broad range of wavelengths, we could obtain the 3D magnetic (from polarimetry) and thermal (from spectroscopy) structure of this atmosphere, scanning from the bottom of the chromosphere (e.g. 
\citealt{busa_chromospheric_1999}) to the top of the transition region and the corona (e.g. \citealt{DelZanna2018a}). This would allow us to infer where hot spots occur in altitude. Importantly, the inferred 3D magnetic field structure in the lower stellar atmosphere can in turn be used in conjunction with the most advanced cool star wind models (e.g. \citealt{chebly_numerical_2023,reville_magnetized_2024}) to derive the 3D planet-star connection (as shown as an example in the right panel Figure 2, see also \citealt{Cohen2015b,Alvarado-Gomez2016,Kavanagh2021,klein_investigating_2024,vidotto_space_2023}, and references therein). The combination of the hot spot localization and magnetic connectivity will allow to (i) unambiguously associate the hot spot to star-planet magnetic interaction and (ii) obtain constraints on the size of the planet magnetosphere (and hence its magnetic field magnitude) from the strength of the hot spot.

\section{Physical Parameters}

\begin{table*}[!ht]
\label{tab:FirstNumbers_direct}
\caption{Physical parameters for the direct detection}
\begin{tabularx}{\textwidth}{CCCCC}
\hline
\textbf{Physical Parameters} & \textbf{State of the Art} & \textbf{Incremental Progress (Enhancing)} & \textbf{Substantial Progress (Enabling)} & \textbf{Major Progress (Breakthrough)}  \\
Number of Objects                                            & -                               & 1                                                                            & 5                                                                           & \textgreater{}10                                                          \\
Spectropolarimetry                                           & Optical+NIR                     & Optical+NIR                                                                  & UV+optical+NIR                                                              & UV+optical+NIR                                                            \\
Distance from Earth (pc)                                     & -                               & \textless{}50                                                                & 100                                                                         & 150  
\\ \hline
\end{tabularx}
\end{table*}

\subsection{Direct measurement of an exoplanet magnetic field requires to quantify the in- and out-of transit linear and circular polarization of the He I 1083 nm triplet}

The polarization of the absorption at the position of the He I 1083 nm triplet is arguably one of the most promising direct detection methods for the magnetic field of an exoplanet. \citet{oklopcic_detecting_2020} showed that for a homogeneous slab atmosphere, the linear polarization in the He I 1083 nm triplet is sensitive to the direction of the local magnetic field for field strengths above a few 10-4 G, and the circular polarisation is sensitive mostly to the line-of-sight magnetic field (Figure \ref{fig:fig1}). These detections require the presence of a large enough population of metastable helium in the exoplanetary atmosphere, which is more likely to be the case for close-in gas giants. The detection threshold also depends on the precision of the polarimetric measurements: 10 G magnetic fields require to measure $Q/I_c$ of about $10^{-4}$. This is also the case for direct detections using the Hanle effect in UV, optical and NIR molecular bands \citep{Berdyugina2006a}, analogous to detections of spatially unresolved 10-15 G fields in the solar atmosphere (e.g., \citealt{berdyugina_evidence_2004}).
The expected levels of polarization are relatively small, reaching a maximum of 0.1\% for favorable geometries \citep{oklopcic_detecting_2020}. By combining multiple subsequent transits (to avoid contamination for longer temporal variabilities), only the brightest targets are supposedly accessible to ground-based spectropolarimeters, such as SPIRou (e.g. WASP-107b; \citealt{allart_high-resolution_2019}--such investigations are currently ongoing). The collecting power of HWO and its stability should allow it to reach small levels of fractional polarization $Q/I$ (at least comparable to ground-based instruments), and make the direct detection viable for fainter targets as well in a single transit. 

The method is intrinsically differential: any detection requires subtracting in- and out-of-transit polarization measurements, thereby naturally removing any constant source of polarization such as one coming from the stellar chromosphere, corona and prominences, where the He I 1083 nm triplet can also form. In addition, the comparison of the helium line with the continuum allows one to remove any broadband source of polarization, such as scattering in the continuum on atoms, molecules or particles (e.g. \citealt{carciofi_polarization_2005,berdyugina_first_2008,berdyugina_polarized_2011,berdyugina_polarized_2016}). Contamination from stellar activity in the transmission polarimetry at 1083 nm is also expected to be negligible (\citealt{cauley_effects_2018}, Mercier et al, 2025 – accepted for pub.)

Therefore, the direct detection of the magnetic field of an exoplanet is based on spectropolarimetric measurements of Stokes $I$, $Q$, $U$ and $V$ at the position of the HeI triplet at 1083 nm carried out during an exoplanetary transit. The large mirror size of HWO enables short exposure times to collect $Q$, $U$ and $V$ separately on the overall short transit timescale. The interpretation and legacy of the observations could be significantly extended by simultaneous, or nearly simultaneous, spectropolarimetric observations at shorter wavelengths, namely in the optical (e.g. H-alpha, CaII H\&K) and ultraviolet (e.g. CII, OI, MgII h\&k), and in UV to NIR molecular bands. These accompanying data would provide context by aiding the interpretation of the signal observed at HeI in terms of stellar activity, for example by helping disentangling exoplanetary and possible stellar activity signals, as well as by supplementing independent evidence for exoplanetary magnetic fields in molecular band polarization. Furthermore, spectropolarimetric observations at shorter wavelengths could lead to identifying additional features in the Stokes profiles that might be connected to exoplanetary magnetic fields, which would enable deepening the study of exoplanetary magnetic fields.

\begin{table*}[!ht]
\label{tab:SecondNumbers_direct}
\caption{Physical parameters for the indirect detection}
\begin{tabularx}{\textwidth}{CCCCC}
\hline
\textbf{Physical Parameters} & \textbf{State of the Art} & \textbf{Incremental Progress (Enhancing)} & \textbf{Substantial Progress (Enabling)} & \textbf{Major Progress (Breakthrough)}  \\
Number of Objects                                            & -                               & 1                                                                            & 5                                                                           & \textgreater{}5                                                           \\
Distance from Earth (pc)                                     & -                               & \textless{}50                                                                & 75                                                                          & 100                                                                       \\
Spectropolarimetry                                           & -                               & FUV                                                                          & FUV+NUV                                                                     & FUV+NUV+optical+NIR                                                      
\\ \hline
\end{tabularx}
\end{table*}

\subsection{Indirect measurement of an exoplanet magnetic field thanks to UV spectropolarimetry}

The thermal and magnetic properties of the atmosphere of a planet-hosting star embed localized hot spots from star-planet magnetic interactions. High-resolution spectroscopy and spectropolarimetry have been carried out in the visible (e.g. with ESPaDONs) and in infra-red (e.g. SPIRou) from ground-based telescopes for a long time \citep{Donati2009,donati_magnetic_2023}, providing a revolutionary view of the photosphere of these stars. The specificity of high-resolution UV spectroscopy is that it probes layers higher up in the atmosphere of the star: the upper chromosphere, the transition region, and the lower corona. The UV range indeed contains multiple spectral lines forming at various temperatures (from about 10 kK to about 10 MK) originating from the chromosphere (e.g. Mg II, CI, OI), the transition region (e.g. CII-IV, NIV, OIII-IV, SiII-IV) and the corona for FUV (e.g. Fe XII). These are the regions where we expect most of the heat deposition from star-planet interactions to occur \citep{paul_stellar_2025}, making UV spectropolarimetry particularly suited to characterize them. 
 
Star-planet magnetic interactions require the planet to be sufficiently close to its host star, i.e. it must orbit within the so-called Alfvén surface of the star (the solar Alfvén surface ranges from about 10 to 20 solar radii; e.g. \citealt{cranmer_suns_2023}). As shown in Figure \ref{fig:fig2}, cooler stars tend to produce wider relative Alfvén surfaces due to stronger large-scale magnetic fields. We expect therefore to have more candidates for magnetic interactions around the coolest stars. These stars also tend to be generally fainter, but also magnetically more active. The collecting power of HWO ($\geq$6 m) will be, in that context, key to be able to detect such signals.     
 
Furthermore, one important limitation of the tentative detections of magnetic interactions so far is the use of chromospheric stellar activity tracers such as Ca II H\&K \citep{Cauley2018}. Such tracers can be hard, but not impossible given sufficient orbital and rotational phase coverage, to disambiguate from intrinsic stellar variability. The combination of tracers originating from various heights in the stellar atmosphere thanks to a NUV-FUV simultaneous coverage would help to isolate the signal coming from a hot spot triggered by the magnetic interaction.   

Finally, the magnetic structuring of the atmosphere of the star, derived from UV spectropolarimetry, can be used in conjunction with 3D stellar wind models to assess the magnetic connectivity between the planet and the star. Such models provide an excellent prediction of the connectivity when compared to observations for the Sun (e.g. \citealt{Riley2011,Lionello2014a,perri_impact_2024}). They have been used also  to model cooler stars (e.g. 
\citealt{Cohen2015b,Folsom2020,Kavanagh2021,chebly_numerical_2023,reville_magnetized_2024}), but challenges still exist as the detailed structuring of the low atmosphere and the associated heating is not fully understood yet. Note that (simultaneously) combining UV spectropolarimetry with visible and/or near infra-red spectropolarimetry from HWO and/or ground-based observatories can, in addition, provide a fully coherent picture of the magnetic properties of the atmosphere of the star.

Therefore, the indirect detection of star-planet interactions is based on the simultaneous measurement of the full Stokes parameters, namely $I$, $Q$, $U$ and $V$, at the position of several UV stellar emission lines tracing different regions of the stellar uppermost layers. The observations would then need to be spaced in time in such a way to enable proper coverage of the exoplanet orbital period (hours-days), stellar rotation period (days-weeks), and stellar activity cycle (months-years).



\section{Description of Observations}

\subsection{Direct measurement of an exoplanet magnetic field requires to quantify the in- and out-of transit linear and circular polarization of the He I 1083 nm triplet}

We envision a survey of transit observations targeting the giant planets for which HeI absorption has already been detected from the ground. Currently, this includes approximately 20 planets (e.g. \citealt{sanz-forcada_connection_2025}). The typical transit duration is of $<$5 hours and an observation should cover as much out-of-transit as in-transit, therefore one transit observation lasts for $<$10 hours. Across this time span, the instrument should be stable enough to enable both polarimetric and high-resolution transmission spectroscopy (details are in the table \ref{tab:FinalNumbers_direct}).

High-spectral resolution ($>$70,000) is required to resolve the HeI lines and possible blends from other stellar spectral features. High spectral resolution also enables measuring any motion of the gas escaping from the planet that provides also information about stellar wind density and velocity. The minimum wavelength range that needs to be covered is that around the HeI metastable triplet, which lies at about 1100 nm. However, the simultaneous observation of additional features probing stellar activity (e.g. Halpha, CaII H\&K, UV emission lines) would provide the means to quantify the possible stellar contamination in the HeI absorption (Stokes I) and polarimetric signal (Stokes Q, U, V). All Stokes parameters should be recorded during the in-transit and out-of-transit phase. The great advantage of a space-based platform for these observations is (1) the lack of telluric contamination and (2) the possibility of simultaneously covering UV stellar lines probing activity over a broad range of altitudes across the upper stellar atmospheric layers, securing the interpretation of the HeI spectropolarimetric observations. The lack of a UV spectropolarimeter either currently available or planned on other facilities makes a possible instrument like Pollux \citep{muslimov_optical_2024} on HWO the only opportunity to carry out this science case.

Figure \ref{fig:fig1} shows that the detection of magnetic fields comparable to those of Earth ($\sim$0.5 G) and Jupiter (polar field $\sim$14 G) requires measuring the level of polarization to the 0.001\% (i.e. $10^{-5}$) level. Such spectropolarimetric measurements require tight wavelength stability, which would then allow to perform high-precision transmission spectroscopy on the Stokes I spectra.

\begin{table*}[!ht]
\label{tab:FinalNumbers_direct}
\caption{Requirements, direct measurements.}
\begin{tabularx}{\textwidth}{CCCCC}
\hline
\textbf{Observation Requirement} & \textbf{State of the Art} & \textbf{Incremental Progress (Enhancing)} & \textbf{Substantial Progress (Enabling)} & \textbf{Major Progress (Breakthrough)}  \\
 \hline
Wavelength range (nm,WENAOKEOA spectropolarimeter)                        & 370-2440                  & 370-2440                                                                              & 270-2440                                                                             & 120-2440                                                                           \\
Stokes parameters                                                          & I, Q, U, V                & I, Q, U, V                                                                            & I, Q, U, V                                                                           & I, Q, U, V                                                                          \\
Spectral resolving power                                                   & 70,000                    & 70,000                                                                                & 100,000                                                                              & \textgreater{}100,000                                                               \\
Telluric contamination                                                     & Yes                       & No                                                                                    & No                                                                                   & No                                                                                  \\
Polarisation level on Stokes Q                                             & 0.01\% / 3\% crosstalk    & 0.01\% (10 G)                                                                         & 0.001\% (1 G; Earth/Jupiter-like)                                                    & 0.001\% (1 G; Earth/Jupiter-like)                                                   \\
Wavelength stability over a transit observation                            & -                         & 10 m/s                                                                                & 1 m/s                                                                                & 1 m/s                                                                              
\\ \hline
\end{tabularx}
\end{table*}

\subsection{Indirect measurement of an exoplanet magnetic field thanks to UV spectropolarimetry}


\begin{table*}[ht!]
    \centering
    \caption[Key targets]{Promising candidates for an indirect detection of exoplanetary magnetic fields. The last candidate in this table, HD 118203b, has an eccentric orbit \citep{castro-gonzalez_signs_2024} and as such is a particularly interesting target for star-planet magnetic interactions (see also \citealt{maggio_coordinated_2015}).}
    \label{tab:targets}
\begin{tabular}{l|llll|llll}
        \noalign{\smallskip}
        \noalign{\smallskip}
Target & $R_P [R_J]$ & $M_P [M_J]$ & $P_{\rm orb}$ [days]  & Spectral type & $R_\star [R_\odot]$ & $M_\star [M_\odot]$ & $P_{\rm rot}^\star$ [days] & G(mag)  \\ \hline
GJ 3138c                    & 0.167                              & 4.18                               & 5.97                                          & K9V                                        & 0.5                                & 0.68                               & 42.5                                 & 10.20                                \\
L 168-9b                    & 0.124                              & 37.35                              & 1.40                                          & M2V                                        & 0.6                                & 0.62                               & 29.8                                 & 10.24                                \\
HD 179949b                  & 1.22                               & 330.51                             & 3.09                                          & F8V                                        & 1.23                               & 1.18                               & 11.0                                 & 6.12                                 \\
HD 189733b                  & 1.14                               & 359.11                             & 2.22                                          & K2V                                        & 0.76                               & 0.92                               & 11.9                                 & 7.43                                 \\
HD 118203b                  & 1.13                               & 689.6                              & 6.1                                           & G0V                                        & 1.993                              & 1.353                              & -                                    & 7.91                
\end{tabular}
\end{table*}

We envision a small survey monitoring the UV full Stokes parameters (I, Q, U, V) of the most promising candidates in terms of star-planet distance and stellar UV flux at Earth, activity, and rotation period. Six of these promising targets are listed in table \ref{tab:targets}. The spectropolarimetric observations shall cover both FUV and NUV simultaneously and should at least cover continuously from the Ly alpha line to the MgII h\&k, and ideally to the CaII H\&K lines. This best ensures coverage of the entire uppermost layers of the stellar atmosphere while the coverage of the CaII lines enables connection with possible ground-based follow-up observations that can be used to intensify and extend the monitoring. Extending the wavelength range further into the red, up to Halpha and the HeI metastable triplet, would strengthen the connection with ground-based observations, and thus aid the interpretation of the space+ground-based observations.

The main lines target of the observations are: Mg II (270.6, 280.3 nm), C I (156.1 nm), O I (130.2, 130.5, 130.6, 135.6 nm), and possibly Ca II (393.4, 396.8 nm) to probe the chromosphere, C II (133.5 nm), C IV (154.8, 155.1 nm), N IV (148.6 nm), O III (166.1, 166.6 nm), O IV (140.2 nm), Si II (126.5, 130.4, 152.7, 153.3 nm), Si III (120.7 nm),  and Si IV (139.4, 140.3 nm) that probe the transition region, and O V (121.8 nm), N V (123.9, 124.3 nm), and Fe XII (124.2 nm) to probe the corona. 

Using the SPI detection from unpolarized light reported by \citet{Cauley2019} as an example, the SPI signal would achieve an S/N greater than 5 when the pixel S/N ranges from 300 to 3000 and the pixel resolution falls within R$\simeq$65,000-110,000. Notably, space-based observations are significantly less affected by Earth atmospheric interference compared to ground-based ones. Therefore, the current HWO instrumental parameters should be sufficient to ensure the proposed signal detection.
The observation sequence should be designed to monitor the star-planet system across multiple relevant timescales, spanning from hours to months or even years. This requirement directly influences the minimum duration of a single observation. The shortest timescale to be probed should account for variations introduced by both the planetary orbital period and the stellar rotation period, which are typically on the order of one day. A random sampling strategy within these periods would be beneficial in reducing the risk of false detections caused by periodic fluctuations and possible under-sampling. On average, we should aim for approximately four observations per day, corresponding to a sampling interval of about six hours, each observation lasting about 15mn thanks to the large collecting area of HWO. In addition, systems with highly eccentric orbits (like HD 118203b, \citealt{castro-gonzalez_signs_2024}, see table \ref{tab:targets}) would be targets for which only specific orbital phases (periastron and apoastron) could be observed, diminishing the total number of visits required for this science case.

The lack of a UV spectropolarimeter either currently available or planned on other facilities makes a possible instrument like Pollux on HWO the only possibility to carry out this science case.

\begin{table*}[!ht]
\label{tab:FinalNumbers_indirect}
\caption{Requirements, indirect measurements.}
\begin{tabularx}{\textwidth}{CCCCC}
\hline
\textbf{Observation Requirement} & \textbf{State of the Art} & \textbf{Incremental Progress (Enhancing)} & \textbf{Substantial Progress (Enabling)} & \textbf{Major Progress (Breakthrough)}  \\
 \hline
Wavelength range (nm; with polarimetry)                                                            & 82-195 (HST COS G140L/800)                        & 123-290                                                                                                       & 123-400                                                                                                      & 920-1100                                                                                                     \\
Stokes parameters                                                                                  & I (COS)                                           & I, Q, U                                                                                                       & I, Q, U, V                                                                                                   & I, Q, U, V                                                                                                  \\
Shortest cadence                                                                                   & every 1.5 hours (HST orbit)                       & every 6 hours                                                                                                 & every 3 hours                                                                                                & every hour                                                                                                   \\
Spectral resolving power                                                                           & $\sim$2,000 (HST COS G140L/800)                   & 10,000                                                                                                        & 50,000                                                                                                       & \textgreater{}100,000                                                                                        
\\ \hline
\end{tabularx}
\end{table*}

{\bf Acknowledgements.} AS and AGM acknowledge funding from the Programme National
de Planétologie (INSU/PNP). AS acknowledges funding from the European
Union’s Horizon-2020 research and innovation programme (grant agreement
no. 776403 ExoplANETS-A) and the PLATO/CNES grant at CEA/IRFU/DAp. AS, CG, AP, FY and the European Research Council project ExoMagnets (grant agreement no.
101125367). DRL would like to acknowledge that this publication has emanated from research conducted with the financial support of Taighde {\'E}ireann – Research Ireland under Grant number 21/PATH-S/9339.  


\end{document}